# Emergence of strong room-temperature ferroelectricity and multiferroicity in 2D-Ti$_3$C$_2$T$_x$ free-standing MXene film

Rabia Tahir, Syedah Afsheen Zahra, Usman Naeem, Syed Rizwan*

Physics Characterization and Simulations Lab (PCSL), Department of Physics, School of Natural Sciences, National University of Sciences and Technology (NUST), H-12 Islamabad, Pakistan

**Corresponding author:** Syed Rizwan; Email: syedrizwan@sns.nust.edu.pk, syedrizwanh83@gmail.com

**Abstract**

Two-dimensional (2D) multiferroics are key candidate materials towards advancement of smart technology. Here, we employed a simple synthesis approach to address the long-awaited dream of developing ferroelectric and multiferroic 2D materials, specially in the new class of materials called MXenes. The etched Ti$_3$C$_2$T$_x$ MXene was first synthesized after HF-treatment followed by delamination process for successful synthesis of free-standing Ti$_3$C$_2$T$_x$ film. The free-standing film was then exposed to air at room-temperature as well as heated at different temperature to form TiO$_2$ layer derived from the Ti$_3$C$_2$T$_x$ MXene itself. TiO$_2$ is reported to be an incipient ferroelectric that assumes a ferroelectric phase in composite form. The structural and morphological analysis confirmed successful synthesis of free-standing film and the Raman spectroscopy revealed the formation of different phases of TiO$_2$. The ferroelectric measurement showed a clear polarization hysteresis loop at room-temperature. Also, the magnetic hysteresis was observed in the film indicating a possibility of coupling between ferroelectric and ferromagnetic phases at room-temperature. The magnetoelectric coupling test was also performed that showed a clear, switchable spontaneous polarization under applied magnetic field. This is the first report on existence of ferroelectric phase and multiferroic coupling in 2D free-standing MXene film at room-temperature which opens-up possibility of 2D materials-based electric and magnetic data storage applications at room-temperature.

**Introduction**

Due to interesting physical and electronic properties for diverse applications, designing multiferroic materials has been a dream for researchers [1-3]. In multiferroics, two or more ferroic phases are coupled through piezoelectric, magnetoelastic or magnetoelectric (ME) interaction in such a way that one ferroic phase is tuned and controlled by another upon external perturbation and vice versa. In ME interaction, the magnetic field (or electric field) can tune the spontaneous polarization (or magnetization) in single material or composite thus, providing a unique opportunity to develop smart sensors, non-volatile memories, small electronic devices, etc that might lead to the miniaturization of electronic gadgets [4]. There are numerous reports on existence of multiferroic phases in bulk materials however, the miniaturization of electronic devices requires developing multiferroicity in materials in low-dimensions [5]. One possible solution is to develop two-dimensional (2D) materials that are smaller in size and easy to synthesize however to the best of our knowledge, the existence of two ferroic phases in 2D materials has not been realized yet.

Ferroelectric materials possess an inherent switchable spontaneous polarization that can be tuned using an external electric field [6]. A very few 2D materials have been predicted to be ferroelectric and the experimental reports on 2D ferroelectrics are even rare. Molybdenum di-sulfide ($MoS_2$), a candidate 2D material from transition metal chalcogenides (TMCs) family, is reported to reveal ferroelectricity under high pressure [7]. The authors created bi-domain polarization under mechanical stress leading to polar distortion in symmetric monolayers generating ferroelectric-type distortion. The binary oxides such as $TiO_2$ are also important material candidates owing to potential applications in dielectric, photocatalysis, etc [8]. $TiO_2$ is also known to be an incipient ferroelectric material that shows ferroelectricity in proximity in its composite form [9]. Y. Yu et al reported successful generation of local broken symmetry causing observation of ferroelectricity in $TiO_2$ [10]. Theoretical predictions also suggest that under high mechanical or uniaxial strain, $TiO_2$ can reveal ferroelectricity at room-temperature however, this is still to be observed experimentally [9, 11].

Transition metal carbides, known as MXenes with general chemical formula $M_{n+1}X_nT_x$, (where, M= Sc, Ti, V, Cr, Zr, Nb, Mo, Hf, Ta and X= C, N while n = 1,2,3), are emerging 2D materials owing to their rich physical and chemical properties suitable for vast avenue of applications such as in transparent conductors [12-13], supercapacitors [14-16], transistors

[17-19], electromagnetic shielding [20], photocatalysis [21], etc. Sunaina et al reported presence of ferromagnetism in $Ti_3C_2T_x$ MXene at room-temperature which was further enhanced upon doping with a magnetic element [22]. Similar studies were also reported in this compound suggesting that $Ti_3C_2T_x$ is a good material candidate among MXene family that might be suitable for spintronic applications [23]. Another report on $Nb_2C$ MXene reported the diamagnetic-type superconductivity with relatively higher transition temperature of 12.5 K [24]. These reports clearly reveal the hidden potential of MXene in magnetic data storage devices, magnetic sensors, etc. Moreover, J. X. Low reported synthesis of $Ti_3C_2$-MXene/$TiO_2$ composite formed by calcination at different temperatures which provides a scheme way to prepare $Ti_3C_2$-MXene composites useful for various applications [25].

In this report, we have employed a simple and cost-effective strategy to develop ferroelectricity in free-standing $Ti_3C_2T_x$ MXene film. Since $TiO_2$ is reported to be an incipient ferroelectric, we assumed the emergence of ferroelectricity in MXene-derived $TiO_2$/$Ti_3C_2T_x$ free-standing film. To prepare this composite with an optimal $TiO_2$ ration in the composite, MXene was air-exposed at room-temperature and was heated at higher temperatures. The presence of $TiO_2$ was confirmed using Raman spectroscopy technique. The ferroelectric testing performed at room-temperature clearly showed the ferroelectric hysteresis. Further, the magnetic testing of the free-standing composite film confirmed the existence of ferromagnetism in the compound thus, paving the way to create the multiferroicity in prepared composite. To confirm this assumption, magnetoelectric testing was also performed that showed a clear magnetic-field control of spontaneous polarization suggesting it to be a very good multiferroic 2D material. Our results, presented here, are the first report on existence of ferroelectricity and multiferroicity induced in MXene-derived $TiO_2$/$Ti_3C_2T_x$ free-standing film at room-temperature prepared using a simple synthesis approach which will open new avenues for 2D MXene multiferroics to be employed in electric and magnetic data storage applications.

**Experimentation**

$Ti_3C_2T_x$ MXene was prepared by selectively eliminating 'Al' from the $Ti_3AlC_2$ MAX phase. The commercial grade $Ti_3AlC_2$ MAX powder was used. In a typical process, 1.0 g of sieved $Ti_3AlC_2$ MAX phase is immersed in the solution containing Hydrofluoric acid (≥ 48 wt %, 1ml HF)

(Sigma Aldrich), Deionized water (3 ml $H_2O$) and Hydrochloric acid (37 wt %, (12M) 6ml HCl) in a ratio of 1:3:6 (volume ratio) under constant stirring at 450 rpm for 24 hours at 35°C. The sieved $Ti_3AlC_2$ powder was steadily added in increments over 5-8 min in a Teflon-lined reaction vessel before it was sealed. Once done, it was rinsed with DI multiple times. After each cycle, the supernatant was discarded to eliminate any undesirable adsorbed ions. The sediment was dispersed in DI until the pH reached 7. The etched powder was then collected by filtration and dried overnight at room temperature.

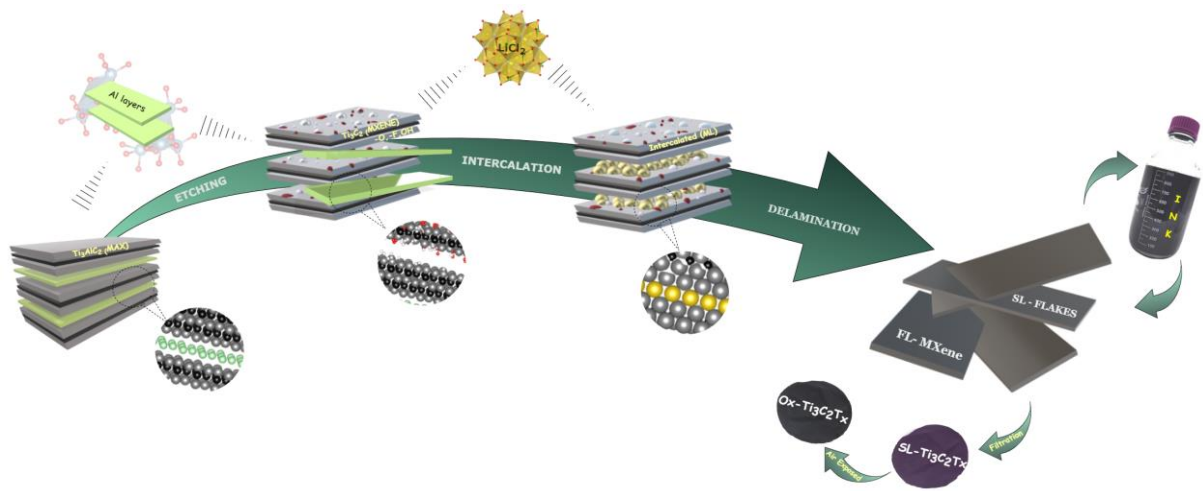

*Figure 1: Synthesis scheme for production of de-laminated $Ti_3C_2T_x$ MXene film.*

Subsequently, the multilayer (ML)-$Ti_3C_2T_x$ MXene was delaminated using LiCl (lithium chloride, 99%) [26]. The ML-MXene powder was dispersed in 20 ml of DI, 1g of LiCl was then added, and the mixture was shaken hard for 8-10 min. The solution was then kept under constant stirring at 300 rpm for 24 hrs. The resulting intercalated dispersion was centrifuged with DI at 3500 rpm for 5 min; the centrifugation was performed repeatedly until a stable colloidal MXene solution was achieved. Following separation, additional DI was added to the sediment, which was then re-dispersed and centrifuged again. This was repeated until the supernatant produced was dilute. The resulting wet sediment formed a clay-like paste which was centrifuged for 30 min at 3500 rpm to ensure that no ML powder remained. The supernatant MXene solution that remained stable after extended centrifugation was vacuum filtered via Celguard membranes, which were then used to form free-standing films. For oxidation, it was reported that $Ti_3C_2T_x$ gets optimum oxidation by heat treatment [27]. So, the synthesized $Ti_3C_2T_x$ MXene film was heat-treated for 2 hours at 100°C (namely

HT@100 °C), 150 °C (namely HT@150 °C), and 300 °C (namely HT@300 °C) under an ambient environment.

**Result and Discussion**

Crystallographic studies of received MAX and synthesized MXene were carried out through X-ray diffraction (XRD) analysis (**Figure 2a**). The XRD pattern of $Ti_3AlC_2$ MAX phase is in good agreement with the literature [28]. As can be seen from XRD pattern, our MAX phase is free of impurities. After the acidic etching and delamination, as expected, (00l) preferential symmetry of basal planes is observed in MXene flakes (indigo). All the peaks beyond 2θ =30° are absent in contrast to MAX phase. Also, the higher order (00l) peaks are of very weak intensity. A significant downshifting of (002) peak from 2θ=9.6° to 2θ= 5.4° is a clear indication of complete exfoliation of MXene via removal of Aluminum layers [29]. This downshift towards a lower angle resulted in an increase in the c-lattice parameter from 18.4Å of MAX to 32.8Å in MXene and an increased d- spacing from 9.2Å to 16.4Å. The oxidation reflections are not visible in the XRD pattern which is due to the low fraction and smaller size of $TiO_2$ crystallites [30] however, the oxidation of MXene under room-temperature is a well-known phenomenon [31-32]. **Figure 2b** exhibits the scanning electron micrograph (SEM) image of the bulk MAX phase while **Figure 2c** is the cross-sectional image of delaminated $Ti_3C_2T_x$ MXene. The morphological image clearly suggests the successful formation of de-laminated MXene with well separated sheets.

Structural and phase transformations were studies through analysis of Raman spectra of the air exposed films heated at 100 °C and 300 °C (**Figure 3**). The peaks present at 409 cm$^{-1}$ and 514 cm$^{-1}$ are ascribed to $TiO_2$ anatase phase [31-32]. The peaks at 154 cm$^{-1}$,

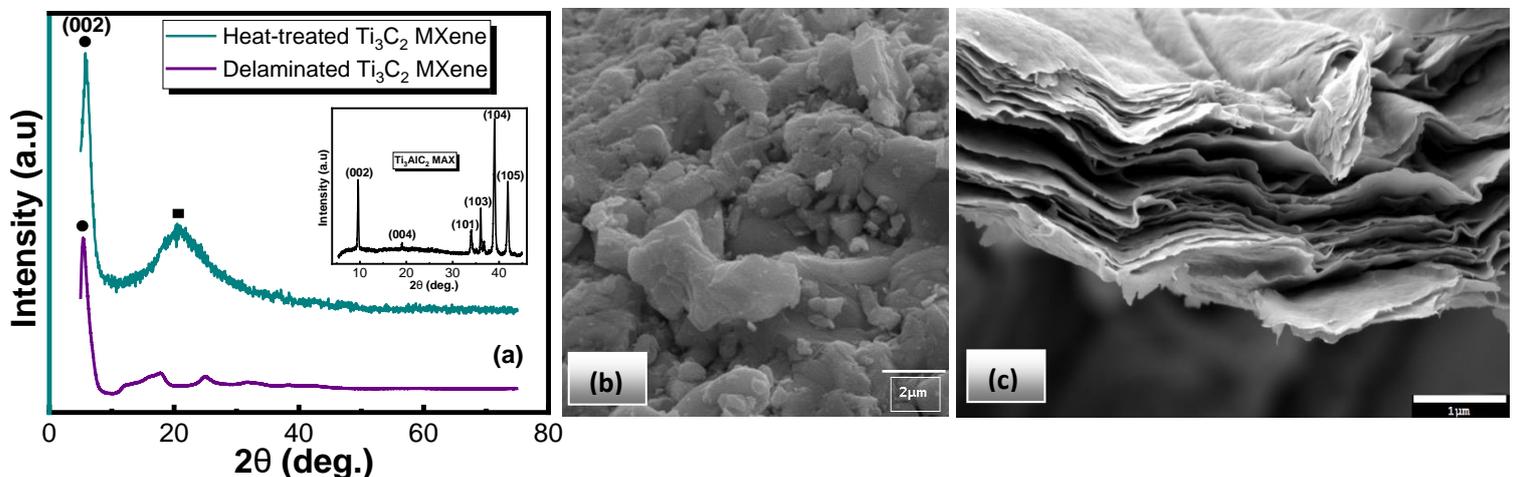

*Figure 2: (a) XRD pattern of $Ti_3C_2T_x$ MXene (**purple**) and heat-treated $Ti_3C_2T_x$ MXene film (**green**); inset: XRD of $Ti_3AlC_2$ MAX phase, (b) SEM micrographs of $Ti_3AlC_2$ MAX and (c) De-laminated $Ti_3C_2T_x$ MXene film.*

260 cm$^{-1}$ and 604 cm$^{-1}$ confirm the presence of rutile phase [33]. Formation of anatase and rutile phases is due to the partial oxidation of Ti$_3$C$_2$T$_x$ MXene over the air exposure and heating effects. Moreover, a long time air exposure as well as heating at high temperature resulted in emergence of another phase, i.e. brookite phase of TiO$_2$. The peaks at 197 cm$^{-1}$ and 409 cm$^{-1}$ in ambient oxidized and 300 °C heat-treated systems are attributed to the brookite phase of TiO$_2$. These modes were not observed in the sample heated at relatively lower temperature (around 100°C) [34–37]. When heated at high temperature, two broader peaks at ≈1390 cm$^{-1}$ and ≈1590 cm$^{-1}$ are characterized as G and D bands of graphitic carbon. Broadening of D and G bands is an evidence of disordered or oxidized carbon, similar to what has been reported for graphene [38]. Presence of TiO$_2$ on the carbon sheets implies that the Ti atoms present in the inner MXene sheets have migrated to the top surface to react with oxygen leaving behind the defects. This kind of outward migration of Ti atoms is similar to the previous studies carried out on TiN [39-40].

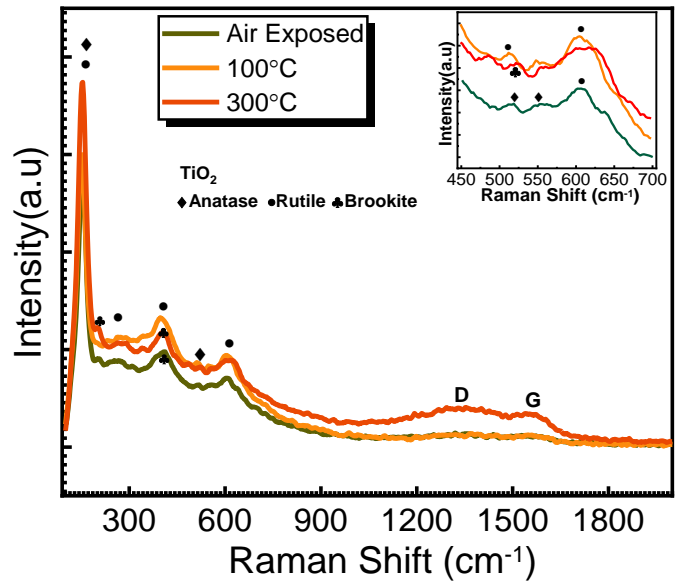

*Figure 3: Raman Spectra of oxidized air-exposed (dark green) and heat-treated samples at 100 $^0$C (orange) and 300 $^0$C Ti$_3$C$_2$T$_x$ MXene film.*

**Figure 4** presents unique ferroelectric, and multiferroic results of the heat-treated de-laminated free-standing Ti$_3$C$_2$T$_x$ MXene film. **Figure 4a** shows spontaneous polarization vs. electric-field hysteresis loops for samples heated at 100 $^0$C (left vertical column) and 150 $^0$C (right vertical column) measured at room-temperature that show clear ferroelectric hysteresis effect. As the electric-field is increased, the polarization increases as well due to alignment of the electric dipoles along the electric-field and reaches to the saturation value

(maximum polarization) of 1 µC/cm$^2$ and 0.6 µC/cm$^2$ for HT@100 $^0$C and HT@150 $^0$C, respectively at +4 kV/cm. When the electric field is reduced to zero, the polarization falls to a non-zero remnant polarization value of 0.5 µC/cm$^2$ and 0.28 µC/cm$^2$ respectively for HT@100 $^0$C and HT@150 $^0$C which is a typical signature of hysteresis effect. It is important to note that the polarization-voltage measurements were also performed for air-exposed sample as well as the ones heated at temperatures higher than 150 $^0$C (**Figure S1**) but the ferroelectric hysteresis effect was not observed. The polarization-voltage curve showed typical conductive behavior which could either be due to the under-formed or over-formed TiO$_2$ phase. This concludes that the induced ferroelectric effect remains stable for a narrow temperature range forming a suitable TiO$_2$/Ti$_3$C$_2$T$_x$ composite ratio. The **inset of Fig. 4a** shows current-voltage (I-V) measurements performed at room-temperature. The current is slightly asymmetric being higher for negative polarity and lower for positive polarity. The I-V results present a typical polarization-switching induced leakage current which is a characteristic of the ferroelectric material [41-42]. **Figure 4b** shows the butterfly-like dielectric constant-voltage hysteresis loops measured for HT@100 $^0$C at room-temperature. It is to be noted that the dielectric constant data is plotted for 1000 cycles that shows its high stability. The dielectric constant is initially small due to the small poling but increases at higher voltage and reaches the maximum at around the coercive voltages. At around the saturation values of polarization, it [43]. **Fig. 4c** represents applied voltage versus time (red) and polarization versus time (orange) measurements. The voltage follows linear change with time in positive and negative polarity which polarizes the electric dipoles along positive and negative directions, respectively indicating fine tuning of the polarization. Also, the polarization curve at minimum (0 sec) and maximum (100 sec) time shows non-zero values which is due to the remanent effect. To the best of our knowledge, this is the first experimental report on existence of ferroelectricity in 2D MXene especially in the de-laminated free-standing film at room-temperature. One possibility for observation of this ferroelectricity in heat-treated MXene is the emergence of TiO$_2$ phase after heat treatment within Ti$_3$C$_2$T$_x$ MXene. Montanari et al [44] performed density function calculations on rutile TiO$_2$ and concluded that it can show ferroelectricity under a uniaxial strain or a negative pressure. Yong Liu et al also predicted a negative pressure induced ferroelectric phase in TiO$_2$ [9]. A. Grunebohm et al computationally the studied effect of strain induced ferroelectricity in TiO$_2$ and concluded that even a non-uniaxial strain could induce a ferroelectric phase in $TiO_2$ where a small change in strain could produce a large spontaneous polarization [11]. The bond length between Ti-O is also reported to be a key parameter in determining proximity for inducing ferroelectricity where a small perturbation could drive the system ferroelectric [45-47].

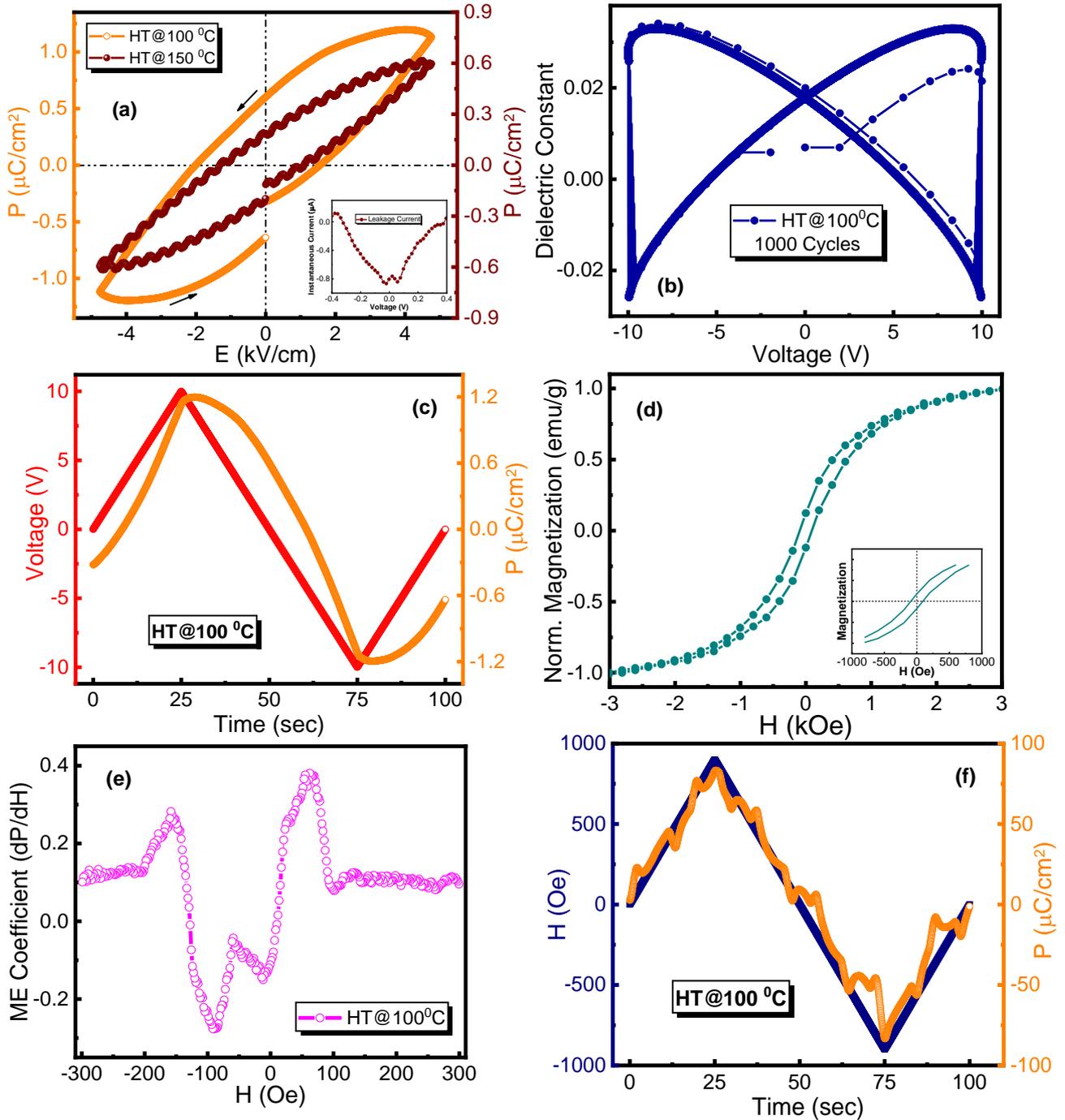

*Figure 4: (a) Ferroelectric hysteresis loops measured for heat-treated de-laminated $Ti_3C_2T_x$ MXene film at 100 $^0C$ (left vertical column) & 150 $^0C$ (right vertical column); inset shows typical leakage current vs voltage that corresponds to the ferroelectric materials, (b) typical dielectric constant as a function of applied voltage for heat-treated sample at 100 $^0C$; the data is presented for 1000 cycles showing its reproducibility and stability, (c) applied voltage (left vertical column) and electric polarization (right vertical column) versus time for heat-treated sample at 100 $^0C$, (d) ferromagnetic hysteresis loop measured at room-temperature; inset shows the near-zero region showing clear magnetic remanence, (e) the magnetoelectric coupling coefficient versus magnetic-field measured for heat-treated sample at 100 $^0C$, (f) applied magnetic-field and spontaneous polarization versus time indicating successful application of magnetic-field onto the sample and clear control of field over the polarization.*

In order to observe the multiferroic effect in $Ti_3C_2T_x$ MXene film, the material must possess at least two ferroic orders (ferromagnetic, ferroelectric, etc) simultaneously. In an earlier literature, Sunaina et al reported the existence of ferromagnetism in $Ti_3C_2T_x$ MXene at room-temperature [22]. Since, our MXene showed a good ferroelectric response, it could show a multiferroic behavior as well if it inherently possesses a ferromagnetic phase. In order to confirm this, we measured magnetization vs. magnetic field loop of the de-laminated $Ti_3C_2T_x$ MXene at room-temperature using vibrating sample magnetometer. The result is shown in **Figure 4d**. It can be seen the results show a good soft ferromagnetic response. The magnetization increases with increase in magnetic field, reaches to its saturation and remanence values at maximum and zero magnetic fields, respectively which is a typical characteristic of a ferromagnetic origin. The **inset of Figure 4d** shows the clear non-zero magnetic remanence in the absence of magnetic field. The results presented in Figure 4(a-d) clearly suggest that our heat-treated $Ti_3C_2T_x$ MXene film could behave like a multiferroic 2D material which can be confirmed by ME coupling effect.

**Figure 4(e-f)** shows magnetoelectric results of heat-treated $Ti_3C_2T_x$ free-standing film measured at room-temperature using multiferroic/ferroelectric precision tester from Radiant Technologies Inc. **Figure 4e** shows the magnetoelectric coupling coefficient versus applied magnetic field measured at room-temperature. The ME coefficient displays obvious peaks at specific magnetic field values which are close to the magnetic coercivity of the material itself (inset: Fig. 4d). This suggests that the magnetic domains are strongly coupled with the electric domains which are well-tuned by the external magnetic field at room-temperature. Thus, our heat-treated $Ti_3C_2T_x$ film proved to possess strong multiferroic property due to the existence of ferroelectric as well as ferromagnetic orders. It is pertinent to mention that our report is the first experimental evidence on existence of multiferroicity in MXene, especially in the 2D de-laminated free-standing $Ti_3C_2T_x$ MXene. Further studies on this effect in layered materials could enhance practical application scope of MXene in electric and magnetic data storage devices in the fields of spintronics and MXetronics.

## Conclusion

We employed a simple approach to induce ferroelectricity and multiferroicity in free-standing heat-treated $Ti_3C_2T_x$ MXene film at room-temperature. The successful synthesized free-standing $Ti_3C_2T_x$ MXene, the films were heated at different temperatures produce $TiO_2$ layer on top of the $Ti_3C_2T_x$ MXene. Raman spectroscopy confirmed the presence of rutile and anatase phases of $TiO_2$. The composite film was tested for ferroelectric measurement that showed a clear hysteresis effect. The effect was attributed to proximity-induced ferroelectricity in $TiO_2$ which was predicted before. Since, our $Ti_3C_2T_x$ MXene film also showed soft ferromagnetic behavior as well so we assumed that the composite film may also possess multiferroic property. To verify this assumption, the magnetoelectric (ME) testing was performed at room-temperature. Interestingly, the film showed a strong ME coupling with a good control of spontaneous polarization via external magnetic field revealing the emergence of strong multiferroic effect as well. Our report is first experimental report on observation ferroelectric as well as multiferroic property in de-laminated heat-treated $Ti_3C_2T_x$ MXene film at room-temperature and presents MXene as the potential 2D material candidate for future data storage devices.

## Authors contribution

Rabia Tahir performed testing and analysis of results, Syedah Afsheen Zahra prepared MXene paper, Usman Naeem helped in sample preparation and Syed Rizwan conceived the idea, performed the measurement and analysis.

## Acknowledgement

The authors thank the Higher Education Commission (HEC) of Pakistan for providing research funding under the Project No.: 20-14784/NRPU/R&D/HEC/2021.

## Data Availability

The data will be available on demand.

## References

[1] Cheong, Sang-Wook, and Maxim Mostovoy. "Multiferroics: a magnetic twist for ferroelectricity." *Nature materials* 6, no. 1 (2007): 13-20.

[2] Rizwan, Syed, S. Zhang, Y. G. Zhao, and X. F. Han. "Exchange-bias like hysteretic magnetoelectric-coupling of as-grown synthetic antiferromagnetic structures." *Applied Physics Letters* 101, no. 8 (2012): 082414.

[3] Rizwan, Syed, G. Q. Yu, S. Zhang, Y. G. Zhao, and X. F. Han. "Electric-field control of CoFeB/IrMn exchange bias system." *Journal of Applied Physics* 112, no. 6 (2012): 064120.

[4] Spaldin, Nicola A., and Rammamoorthy Ramesh. "Advances in magnetoelectric multiferroics." *Nature materials* 18, no. 3 (2019): 203-212.

[5] Lu, Chengliang, Weijin Hu, Yufeng Tian, and Tom Wu. "Multiferroic oxide thin films and heterostructures." *Applied physics reviews* 2, no. 2 (2015): 021304.

[6] Qiao, Huimin, Chenxi Wang, Woo Seok Choi, Min Hyuk Park, and Yunseok Kim. "Ultra-thin ferroelectrics." *Materials Science and Engineering: R: Reports* 145 (2021): 100622.

[7] Lipatov, Alexey, Pradeep Chaudhary, Zhao Guan, Haidong Lu, Gang Li, Olivier Crégut, Kokou Dodzi Dorkenoo et al. "Direct observation of ferroelectricity in two-dimensional $MoS_2$." *npj 2D Materials and Applications* 6, no. 1 (2022): 1-9.

[8] Yu, Yang, Li-Dong Wang, Wei-Li Li, Yu-Long Qiao, Yu Zhao, Yu Feng, Tian-Dong Zhang, Rui-Xuan Song, and Wei-Dong Fei. "Room temperature ferroelectricity in donor-acceptor co-doped $TiO_2$ ceramics using doping-engineering." *Acta Materialia* 150 (2018): 173-181.

[9] Liu, Yong, Lihong Ni, Zhaohui Ren, Gang Xu, Chenlu Song, and Gaorong Han. "Negative pressure induced ferroelectric phase transition in rutile $TiO_2$." *Journal of Physics: Condensed Matter* 21, no. 27 (2009): 275901.

[10] Yu, Yang, Li-Dong Wang, Wei-Li Li, Yu-Long Qiao, Yu Zhao, Yu Feng, Tian-Dong Zhang, Rui-Xuan Song, and Wei-Dong Fei. "Room temperature ferroelectricity in donor-acceptor co-doped $TiO_2$ ceramics using doping-engineering." *Acta Materialia* 150 (2018): 173-181.

[11] Grünebohm, A., C. Ederer, and P. Entel. "First-principles study of the influence of (110)-oriented strain on the ferroelectric properties of rutile $TiO_2$." *Physical Review B* 84, no. 13 (2011): 132105.

[12] Halim, Joseph, Maria R. Lukatskaya, Kevin M. Cook, Jun Lu, Cole R. Smith, Lars-Åke Näslund, Steven J. May et al. "Transparent conductive two-dimensional titanium carbide epitaxial thin films." *Chemistry of Materials* 26, no. 7 (2014): 2374-2381.

[13] Mariano, Marina, Olha Mashtalir, Francisco Q. Antonio, Won-Hee Ryu, Bingchen Deng, Fengnian Xia, Yury Gogotsi, and André D. Taylor. "Solution-processed titanium carbide MXene films examined as highly transparent conductors." *Nanoscale* 8, no. 36 (2016): 16371-16378.

[14] Lai, Shen, Jaeho Jeon, Sung Kyu Jang, Jiao Xu, Young Jin Choi, Jin-Hong Park, Euyheon Hwang, and Sungjoo Lee. "Surface group modification and carrier transport properties of layered transition metal carbides ($Ti_2CTx$, T:–OH,–F and–O)." *Nanoscale* 7, no. 46 (2015): 19390-19396.

[15] Lukatskaya, Maria R., Olha Mashtalir, Chang E. Ren, Yohan Dall'Agnese, Patrick Rozier, Pierre Louis Taberna, Michael Naguib, Patrice Simon, Michel W. Barsoum, and Yury Gogotsi. "Cation intercalation and high volumetric capacitance of two-dimensional titanium carbide." *Science* 341, no. 6153 (2013): 1502-1505.

[16] Ghidiu, Michael, Maria R. Lukatskaya, Meng-Qiang Zhao, Yury Gogotsi, and Michel W. Barsoum. "Conductive two-dimensional titanium carbide 'clay' with high volumetric capacitance." *Nature* 516, no. 7529 (2014): 78-81.

[17] Yang, Yajie, Sima Umrao, Shen Lai, and Sungjoo Lee. "Large-area highly conductive transparent two-dimensional $Ti_2CT_x$ film." *The journal of physical chemistry letters* 8, no. 4 (2017): 859-865.

[18] Naguib, Michael, Joseph Halim, Jun Lu, Kevin M. Cook, Lars Hultman, Yury Gogotsi, and Michel W. Barsoum. "New two-dimensional niobium and vanadium carbides as promising materials for Li-ion batteries." *Journal of the American Chemical Society* 135, no. 43 (2013): 15966-15969.

[19] Rakhi, Raghavan B., Bilal Ahmed, Mohamed N. Hedhili, Dalaver H. Anjum, and Husam N. Alshareef. "Effect of postetch annealing gas composition on the structural and electrochemical properties of $Ti_2CT_x$ MXene electrodes for supercapacitor applications." *Chemistry of Materials* 27, no. 15 (2015): 5314-5323.

[20] Xie, Yu, Yohan Dall'Agnese, Michael Naguib, Yury Gogotsi, Michel W. Barsoum, Houlong L. Zhuang, and Paul RC Kent. "Prediction and characterization of MXene nanosheet anodes for non-lithium-ion batteries." *ACS nano* 8, no. 9 (2014): 9606-9615.

[21] Liu, Fanfan, Aiguo Zhou, Jinfeng Chen, Heng Zhang, Jianliang Cao, Libo Wang, and Qianku Hu. "Preparation and methane adsorption of two-dimensional carbide $Ti_2C$." *Adsorption* 22, no. 7 (2016): 915-922.

[22] Rafiq, Sunaina, SaifUllah Awan, Ren-Kui Zheng, Zhenchao Wen, Malika Rani, Deji Akinwande, and Syed Rizwan. "Novel room-temperature ferromagnetism in Gd-doped 2-dimensional $Ti_3C_2T_x$ MXene semiconductor for spintronics." *Journal of Magnetism and Magnetic Materials* 497 (2020): 165954.

[23] Iqbal, Mehroz, Jameela Fatheema, Qandeel Noor, Malika Rani, Muhammad Mumtaz, Ren-Kui Zheng, Saleem Ayaz Khan, and Syed Rizwan. "Co-existence of magnetic phases in two-dimensional MXene." *Materials Today Chemistry* 16 (2020): 100271.

[24] Babar, Zaheer Ud Din, M. S. Anwar, Muhammad Mumtaz, Mudassir Iqbal, Ren-Kui Zheng, Deji Akinwande, and Syed Rizwan. "Peculiar magnetic behaviour and Meissner effect in two-dimensional layered $Nb_2C$ MXene." *2D Materials* 7, no. 3 (2020): 035012.

[25] Low, Jingxiang, Liuyang Zhang, Tong Tong, Baojia Shen, and Jiaguo Yu. "$TiO_2$/MXene $Ti_3C_2$ composite with excellent photocatalytic $CO_2$ reduction activity." *Journal of Catalysis* 361 (2018): 255-266.

[26] C.E. Shuck, A. Sarycheva, M. Anayee, A. Levitt, Y. Zhu, S. Uzun, V. Balitskiy, V. Zahorodna, O. Gogotsi, Y. Gogotsi, Scalable synthesis of Ti3C2Tx mxene, Adv. Eng. Mater. 22 (2020) 1901241.

[28] Shuck, Christopher E., Asia Sarycheva, Mark Anayee, Ariana Levitt, Yuanzhe Zhu, Simge Uzun, Vitaliy Balitskiy, Veronika Zahorodna, Oleksiy Gogotsi, and Yury Gogotsi. "Scalable synthesis of $Ti_3C_2T_x$ mxene." *Advanced Engineering Materials* 22, no. 3 (2020): 1901241.

[29] Lipatov, Alexey, Mohamed Alhabeb, Maria R. Lukatskaya, Alex Boson, Yury Gogotsi, and Alexander Sinitskii. "Effect of synthesis on quality, electronic properties and environmental stability of individual monolayer $Ti_3C_2$ MXene flakes." *Advanced Electronic Materials* 2, no. 12 (2016): 1600255.

[30] Chertopalov, Sergii, and Vadym N. Mochalin. "Environment-sensitive photoresponse of spontaneously partially oxidized Ti$_3$C$_2$ MXene thin films." *ACS nano* 12, no. 6 (2018): 6109-6116.

[31] Cao, Minjuan, Fen Wang, Lei Wang, Wenling Wu, Wenjing Lv, and Jianfeng Zhu. "Room temperature oxidation of Ti$_3$C$_2$ MXene for supercapacitor electrodes." *Journal of The Electrochemical Society* 164, no. 14 (2017): A3933.

[32] Zhang, Chuanfang John, Sergio Pinilla, Niall McEvoy, Conor P. Cullen, Babak Anasori, Edmund Long, Sang-Hoon Park et al. "Oxidation stability of colloidal two-dimensional titanium carbides (MXenes)." *Chemistry of Materials* 29, no. 11 (2017): 4848-4856.

[33] Li, Xinliang, Xiaowei Yin, Meikang Han, Changqing Song, Hailong Xu, Zexin Hou, Litong Zhang, and Laifei Cheng. "Ti$_3$C$_2$ MXenes modified with in situ grown carbon nanotubes for enhanced electromagnetic wave absorption properties." *Journal of Materials Chemistry C* 5, no. 16 (2017): 4068-4074.

[34] J. Chen, M. Guan, X. Zhang, X. Gong, Insights into a rutile/brookite homojunction of titanium dioxide: Separated reactive sites and boosted photocatalytic activity, RSC Adv. 9 (2019) 36615–36620.

[35] G.A. Tompsett, G.A. Bowmaker, R.P. Cooney, J.B. Metson, K.A. Rodgers, J.M. Seakins, The Raman spectrum of brookite, TiO2 (Pbca, Z = 8), J. Raman Spectrosc. 26 (1995) 57–62.

[36] R. Verma, J. Gangwar, A.K. Srivastava, Multiphase TiO2 nanostructures: A review of efficient synthesis, growth mechanism, probing capabilities, and applications in bio-safety and health, RSC Adv. 7 (2017).

[37] M.C. Ceballos-Chuc, C.M. Ramos-Castillo, J.J. Alvarado-Gil, G. Oskam, G. Rodríguez-Gattorno, Influence of brookite impurities on the raman spectrum of TiO2 anatase nanocrystals, J. Phys. Chem. C. 122 (2018) 19921–19930.

[38] Y. Yoon, T.A. Le, A.P. Tiwari, I. Kim, M.W. Barsoum, H. Lee, Low temperature solution synthesis of reduced two dimensional Ti3C2 MXenes with paramagnetic behaviour, Nanoscale. 10 (2018) 22429–22438.

[39] Gogotsi, Yu G., and F. Porz. "The oxidation of particulate-reinforced Si$_3$N$_4$-TiN composites." *Corrosion science* 33, no. 4 (1992): 627-640.

[40] Naguib, Michael, Olha Mashtalir, Maria R. Lukatskaya, Boris Dyatkin, Chuanfang Zhang, Volker Presser, Yury Gogotsi, and Michel W. Barsoum. "One-step synthesis of nanocrystalline transition metal oxides on thin sheets of disordered graphitic carbon by oxidation of MXenes." *Chemical communications* 50, no. 56 (2014): 7420-7423.

[41] L. Pintilie, et al, "Ferroelectric polarization-leakage current in high quality epitaxial Pb(Zr, Ti)O3 films." *Phys. Rev. B,* 75 (2007): 104103.

[42] A.G. Boni, et al, "Study of the leakage current in epitaxial ferroelectric Pb(Zr0.52Ti0.48)O3 layer with SrRuO3 bottom electrode and different metals as top contacts." *Dig. J. Nanomater. Biostr.* 10 (2015): 1257.

[43] Kim, S. J., et al, "Ferroelectric TiN/Hf0.5Zr0.5O2/TiN capacitors with low-voltage operation and high reliability for next-generation FRAM applications." *IEEE* 18 (2018): 5386.

[44] Montanari, B., and N. M. Harrison. "Pressure-induced instabilities in bulk TiO2 rutile." *Journal of Physics: Condensed Matter* 16, no. 3 (2004): 273.

[45] Barrett, John H. "Dielectric constant in perovskite type crystals." *Physical Review* 86, no. 1 (1952): 118.

[46] Ang, Chen, Zhi Yu, P. M. Vilarinho, and J. L. Baptista. "B i: $SrTiO_3$: A quantum ferroelectric and a relaxor." *Physical Review B* 57, no. 13 (1998): 7403.

[47] Mitsui, T., and William Blackburn Westphal. "Dielectric and X-Ray Studies of $Ca_xBa_{1-x}Ti_3$ and $Ca_x Sr_{1-x} TiO_3$." *Physical Review* 124, no. 5 (1961): 1354.